\documentclass[twocolumn,superscriptaddress,showpacs,amsfonts]{revtex4-1}
\pdfoutput=1

\usepackage{epsfig}
\usepackage{graphicx}

\begin{document}
\title{Tuning Pairing Amplitude and Spin-Triplet Texture by Curving Superconducting Nanostructures} 
\author{Zu-Jian Ying}
\affiliation{CNR-SPIN and Dipartimento di Fisica ``E. R. Caianiello'',
Universit\`a di Salerno, I-84084 Fisciano (Salerno), Italy}
\affiliation{Beijing Computational Science Research Center, Beijing 100084, China}
\author{Mario Cuoco}
\affiliation{CNR-SPIN and Dipartimento di Fisica ``E. R. Caianiello'',
Universit\`a di Salerno, I-84084 Fisciano (Salerno), Italy}
\author{Carmine Ortix}
\affiliation{Institute for Theoretical Physics, Center for Extreme Matter and Emergent Phenomena, Utrecht University, Princetonplein 5, 3584 CC Utrecht, The Netherlands}
\affiliation{Institute for Theoretical Solid State Physics, IFW-Dresden, Helmholtzstr. 20, D-01069 Dresden, Germany}

\author{Paola Gentile}
\affiliation{CNR-SPIN and Dipartimento di Fisica ``E. R. Caianiello'',
Universit\`a di Salerno, I-84084 Fisciano (Salerno), Italy}

\begin{abstract}
We investigate the nature of the superconducting state in curved 
nanostructures with Rashba spin-orbit coupling (RSOC).
In bent nanostructures with inhomogeneous curvature we find a local enhancement or suppression of the superconducting order parameter, with the effect that can be tailored by tuning either the RSOC strength or the carrier density.
Apart from the local superconducting spin-singlet amplitude control, the geometric curvature generates non-trivial textures of the spin-triplet pairs through a spatial variation of the ${\vec{d}}$-vector. 
By employing the representative case of an elliptically deformed quantum ring, we demonstrate that the amplitude of the ${\vec{d}}$-vector strongly depends on the strength of the local curvature and it generally exhibits a three-dimensional profile whose winding is tied to that of the single electron spin in the normal state.
Our findings unveil novel paths to manipulate the quantum structure of the superconducting state in RSOC nanostructures through their geometry.    
\end{abstract}
\maketitle
\noindent 
{\it{Introduction -- }}
Inversion symmetry breaking is a fundamental property that yields spin-orbit locking and set the structure of both single electron and paired quantum states.
Indeed, on the one hand, for low-dimensional semiconducting nanosystems with structure inversion asymmetry, the Rashba spin-orbit coupling (RSOC)~\cite{Dresselhaus1955,Rashba1960,Bychkov1984} allows to tune the spin orientation through the electron propagation and vice versa to exert a spin control on the electron trajectories. On the other hand, lack of inversion symmetry in superconductors
makes neither spin nor parity good quantum numbers anymore. 
The ensuing mixing of even spin-singlet and odd spin-triplet channels~\cite{Gorkov2001,Frigeri2004} leads to a series of novel features ranging from the anomalous magneto-electric~\cite{Lu2008} 
effect, to unconventional surface states~\cite{Vorontsov2008}, and
topological phases~\cite{Tanaka2009,Sato2009}.
A major boost in the framework of inversion asymmetric systems relied on the proposal~\cite{Lutchyn2010,Oreg2010} and the observation of a topological superconducting phase~\cite{Mourik2012,Deng2012,Das2012,Chiurchill2013}, hosting end Majorana modes, in a one-dimensional semiconductor nanowire with sizable RSCO and proximity-induced superconductivity.
Although the investigation of Majorana modes primarily focuses on single semiconducting 
wires, the path for Majorana detection and braiding often requires to employ more  
complex networks~\cite{Alicea2011,Halperin2012,Clarke2011} or suggests alternative curved geometries,
e.g. circular \cite{Pientka2013,Scharf2015,Rosenstein2013,Lee2014} and elliptical quantum rings~\cite{Ghazaryan2016}.

Studies of RSOC semiconducting rings have opened the path to a geometric design of the electron spin and quantum geometric phase as predicted \cite{Frustaglia2004,Saarikoski2015} and experimentally demonstrated through the application of an external magnetic field \cite{Nagasawa2013}.  
Recently, the theoretical analysis of shape-deformed RSOC nanostructures established a deep connection between electronic spin textures, spin transport properties, and the nanoscale shape of the system, thus providing foundational elements for an all-geometrical and electrical control of the spin orientation~\cite{Ying2016}, including the possibility of topological non-trivial phases~\cite{Gentile2015,Ying2016}.  
Along this line, due to the expected strong impact of inversion symmetry breaking on single and paired electronic states, fundamental questions naturally arise on the character of the superconducting state in RSOC shape deformed superconductors or curved semiconductor-superconductor heterostructures.

In this Letter, we aim to unveil the interplay between shape deformations and superconductivity in RSOC nanostructures. While in systems with constant curvature (e.g. quantum wires
or circular rings) the RSOC is monotonously affecting the superconductivity, the spatial variation of the Rashba field through the curvature of the nanostructure can twist the effective electron mass such as to yield either a local enhancement or a suppression of the superconducting order parameter. 
We demonstrate this effect by employing elliptically shaped quantum rings and we provide evidence for  
its control trough both the RSOC strength and electron density variations.
Apart from driving the superconducting spin-singlet amplitude, the inhomogeneous profile of the curvature generates non-trivial spatial patterns of the spin-triplet pairs via the modulation of the amplitude and orientation of the ${\vec{d}}$-vector. 
We show that the geometric curvature can effectively tailor the spin-triplet pairing by yielding three-dimensional spatial textures, and the behavior of the superconducting ${\vec{d}}$-vector generally follows the evolution of the electron spin-orientation in the normal state. Such findings indicate that the curvature can effectively lead to a spin-torque on the spin of the electron pairs.   

\begin{figure}
\includegraphics[width=0.95\columnwidth]{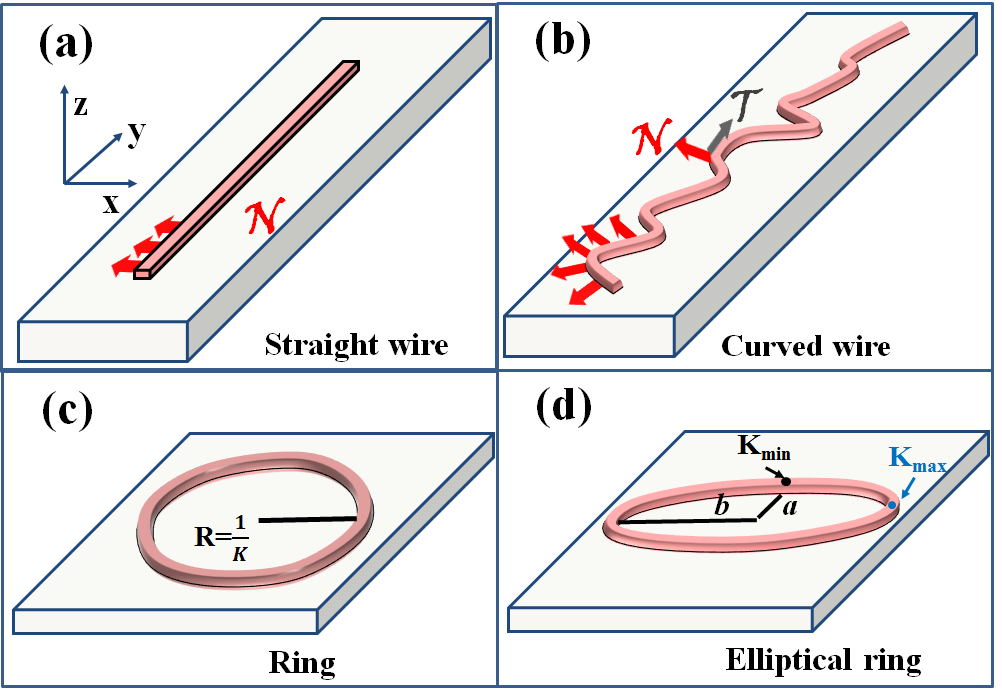}
\caption{(color online). Schematics of the geometric profile of (a) straight, (b) planarly curved nanostructure, (c) ring, and (d) elliptically deformed ring with semi-axes $a$ and $b$. Red (gray) arrows indicate the perpendicular (tangential) direction of the spin orientation with respect to the geometric profile. $K$ is the curvature of the ring, e.g. the inverse of the its radius $R$. Black (blue) dots indicate a position with minimum $K_{min}$ (maximal $K_{max}$) amplitude of the curvature in the elliptical ring, respectively.}
\label{fig:fig1}
\end{figure}

{\it{Model -- }}
We consider electrons moving along a one-dimensional planarly curved nanostructure [Fig. \ref{fig:fig1}] in the presence of a local spin-singlet pairing interaction. 
Due to the structural inversion symmetry breaking the electrons are subject to a RSOC,  
which, as in the case of a not-curved nanostructure [Fig. \ref{fig:fig1}(a)], couples the orbital and the local spin component normal to the electron motion~\cite{Gentile2013}. 
Since the nanostructure has a non trivial geometric profile, the spin-orientation perpendicular to the momentum of the quasiparticle changes its direction and thus it is spatially dependent. Such aspect
can be conveniently expressed by introducing the local normal $\hat{\cal N}(s)$ and tangential $\hat{\cal T}(s)$ directions at a given position $s$ along the given curve, as well as the corresponding local Pauli matrices for the spin components, i.e.
$\sigma_N(s) = \mathbf{\tau} \cdot \hat{\cal N}(s)$ and $\sigma_T(s) = \mathbf{\tau} \cdot \hat{\cal T}(s)$ in the moving frame of the electrons, with $\mathbf{\tau}$ being the usual Pauli matrices [Fig. \ref{fig:fig1}(a)].
Then, after having decoupled the pairing interaction, the Bogoliubov-de Gennes Hamiltonian for a planar non-uniformly curved nanostructure with RSOC and local spin-singlet pairing interaction can be written as ~\cite{Gentile2013,Ortix2015,Gentile2015,Zhang2007,Ying2016}:
\begin{eqnarray*}
{\cal H}&& = \int ds\,c^{\dagger}_{\sigma}(s) \{ (-\frac{\hbar^2}{2 m^{\star}} \partial_s^2 -\mu) + 
\frac{i \alpha_{SO}}{2} [ \sigma_N(s) \partial_s + \nonumber \\
&& \partial_s \sigma_N(s) ] \} c_{\sigma'}(s)  + g [ \Delta(s) c^{\dagger}_{\uparrow}(s) c^{\dagger}_{\downarrow}(s)  + h.c. ]+\frac{|\Delta(s)|^2}{g},
\label{eq:hamiltoniankp}
\end{eqnarray*}
where $c^{\dagger}_{\sigma}(s)$ creates an electron with spin $\sigma$ at the position $s$, being the arclength of the planar curve measured from an arbitrary reference point, $m^{\star}$ is the effective mass of the charge carriers, $\alpha_{SO}$ is the RSOC, $g$ is the interaction in the spin-singlet channel, and $\mu$ stands for the chemical potential. The amplitude $\Delta(s)=\langle c_{\uparrow}(s) c_{\downarrow}(s) \rangle$ indicates the expectation value on the ground-state of the local spin-singlet pairing correlator, and it is determined self-consistently until the minimum of the free energy is achieved within the desired accuracy. 
Eq. \ref{eq:hamiltoniankp} generalizes the Hamiltonian originally proposed for a quantum ring with constant curvature ~\cite{Meijer2002} and includes an inhomogeneous curvature profile, as well as
the possibility of having a local $s$-wave pairing coupling.
For a general bent nanowire [Fig. 1 (b))], the normal and tangential directions to the curve can be expressed in terms of a polar angle $f(s)$ as $\hat{\cal N}(s) = \left\{\cos{f(s)},\sin{f(s)} ,0 \right\}$, and  $\hat{\cal T}(s) = \left\{\sin{f(s)} , -\cos{f(s)},0 \right\}$. The polar angle is related to the local curvature $K(s)$ via $\partial_s f(s) = -K(s)$. To compute the spatial dependent pairing amplitude we follow the conventional mapping of the Hamiltonian from continuum-to-lattice \cite{Gentile2015,Ying2016} and solve the inhomogeneous Bogoliubov-de Gennes equations in order to evaluate the local superconducting order parameter (OP) amplitude \cite{Cuoco2008}. The analysis has been performed for systems size up to $L=400$ sites. Greater values do not change qualitatively the results.   
\begin{figure}
\includegraphics[width=0.95\columnwidth]{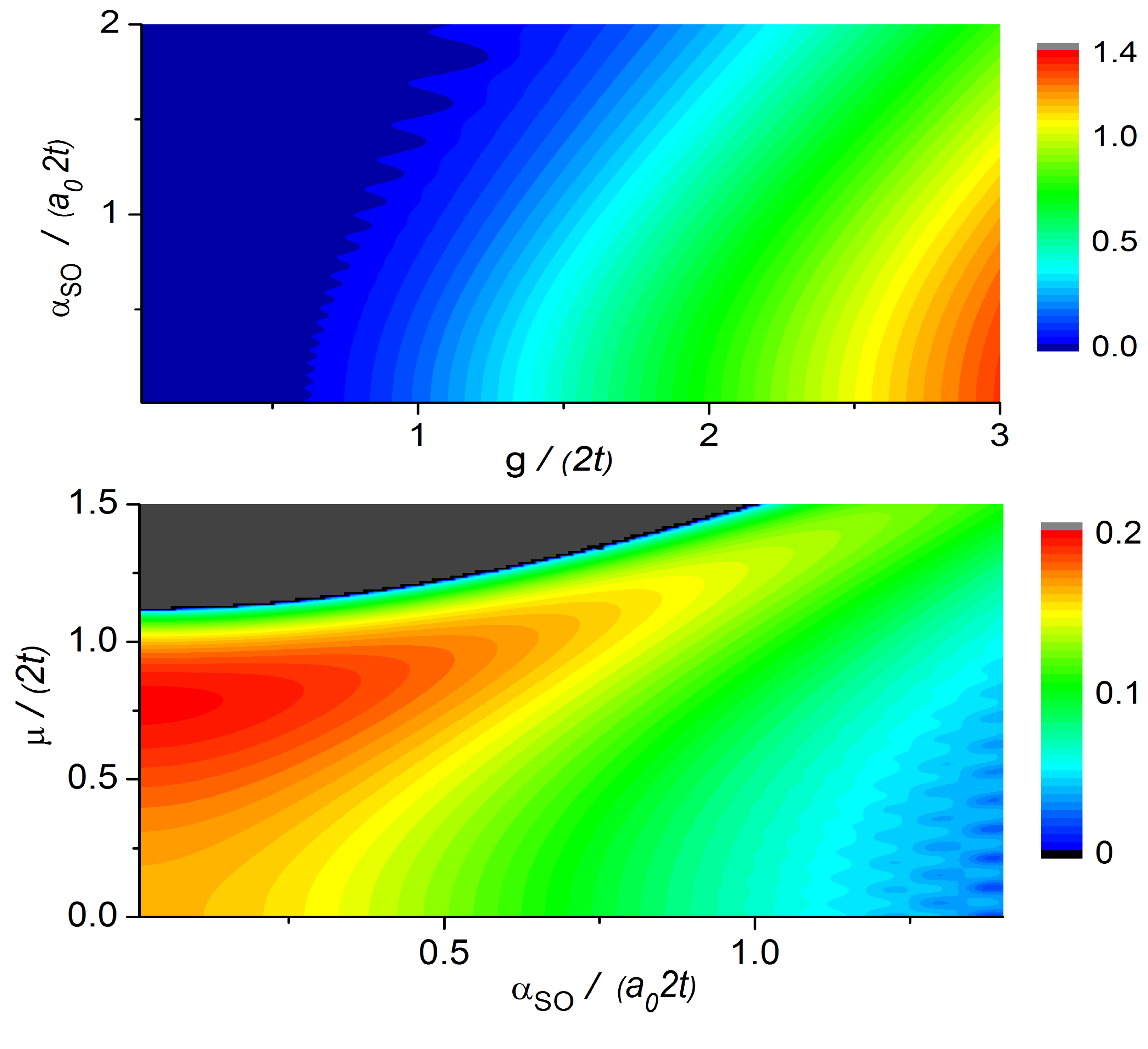}
\caption{(color  online). Contour map of the averaged amplitude of the spin-singlet OP by varying the electron density via the chemical potential $\mu$, the RSOC $\alpha_{SO}$ and the superconducting pairing coupling $g$. $a_0$ is the inter-atomic distance and $t$ the nearest-neighbor hopping amplitude in the effective tight-binding model. In (a) $\mu=0.$ (half-filling) and $g/2t$=1.0 in (b). }
\label{fig:fig2}
\end{figure}
\begin{figure}
\includegraphics[width=0.95\columnwidth]{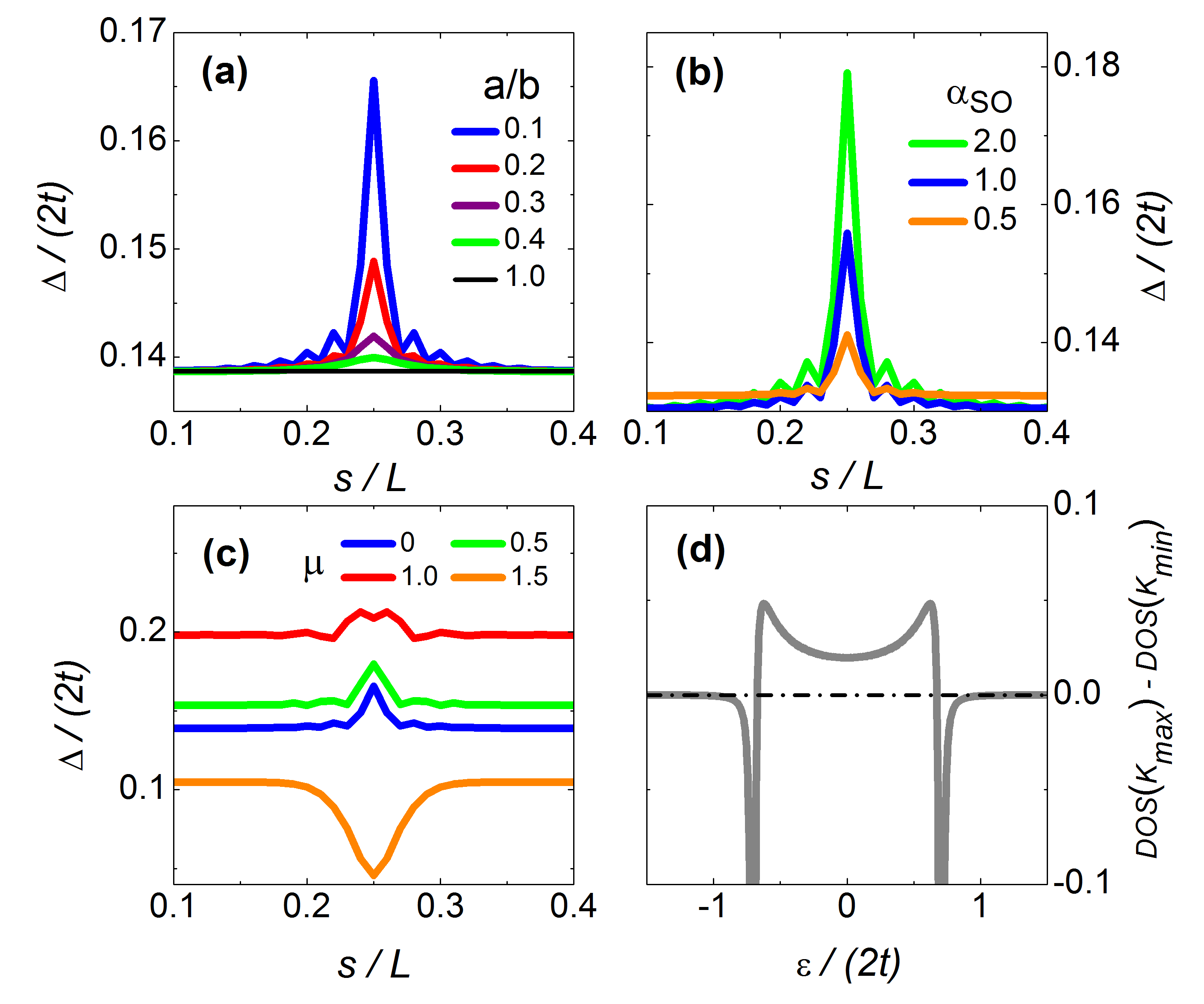}
\caption{(color  online). Evolution of the spin-singlet order parameter close to the position of maximal curvature (i.e. $s/L=0.25$) in the elliptical ring as a function of (a) the elliptical squeezing ratio $a/b$, (b) the RSOC strength, (c) the chemical potential. (d) Difference between the local density of states at the position of maximal ($K_{max}$) and minimal ($K_{min}$) curvature of the elliptical ring (see Fig. 1). For convenience and clarity of graphical representation we employed the following sets of microscopic parameters: (a) $\mu=0.0$, $\alpha_{SO}=1.0$, $g=1.2$, (b) $a/b=0.1$, $\mu=0.0$, $g=1.0,1.18,1.66$ for $\alpha_{SO}=0.5,1.0,0.5$, respectively. In (c) $\alpha_{SO}=1.0$, $a/b=0.1$, $g=1.2$. For the panel (d) we have $\alpha_{SO}=1.0$, $a/b=0.1$, $g=1.0$. All the energy scales are in unit of $2t$.}
\label{fig:fig2}
\end{figure}
{\it{Phase diagram --}}
We start by considering how the superconducting state evolves by varying the RSOC strength, the electron density, and the shape of the nanostructures by comparing the cases of a quantum wire, a circular, and an elliptical ring. The aim is to establish which role the shape plays in tailoring the amplitude of the superconducting OP. 
Remarkably, we find a sort of universal behavior for the averaged OP over the length of the system, when comparing the results for the three geometries.
The spatially averaged OP does not exhibit qualitative fingerprints allowing to single out the different shapes of the RSOC nanostructures. Indeed, the general tendency is to have a detrimental impact on the superconducting state by increasing the RSOC with a critical threshold above which a significant suppression of superconductivity occurs, as signaled by the strong reduction of the OP.
The threshold is dependent on the superconducting coupling $g$, such as larger values of $\alpha_{SO}$ are needed to suppress the superconducting state when moving from a weakly to a strongly coupled regime [Fig. 2]. The phase diagram for different filling concentrations indicates that the most favorable conditions for the superconductivity are obtained when the Fermi level is close to the band edge (i.e. $\mu \sim 2 t$), consistently with the position of the maximum of the density of states due to the one-dimensional hopping connectivity in all the examined systems. When the RSOC becomes comparable with the spin independent hopping $t$, the superconducting state can survive in a broader region of effective chemical potentials as a consequence of the bandwidth enlargement driven by the RSOC [Fig. 2]. 

\begin{figure*}[t!]
\includegraphics[height=5 cm, width=0.99\textwidth]{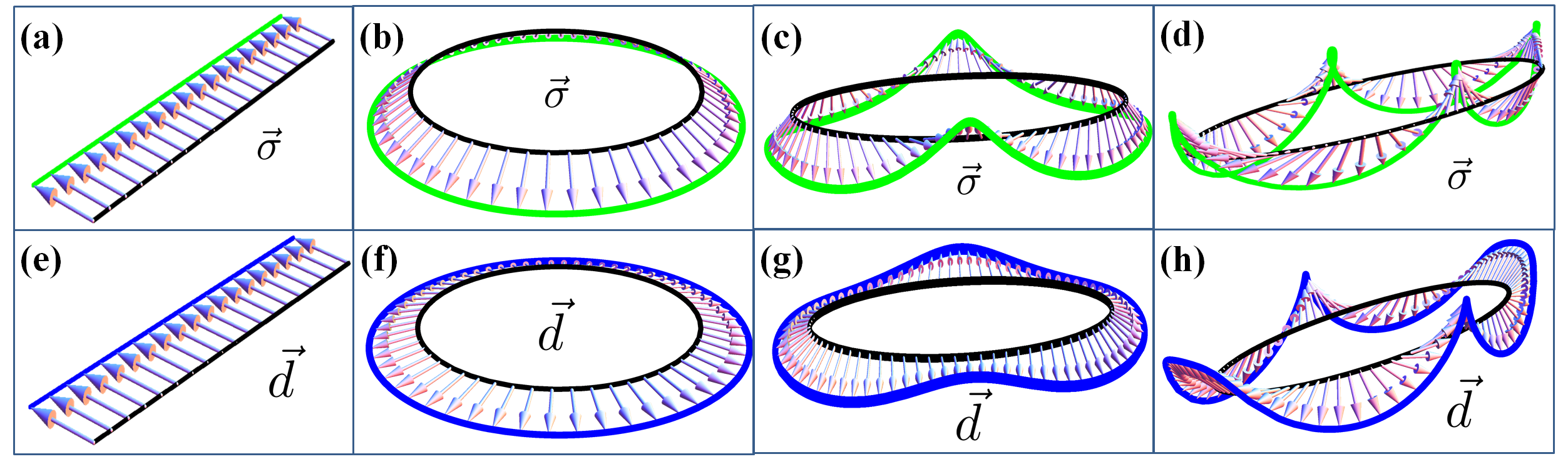}
\caption{(color  online). Representative configurations of the spatial profile of the electron spin orientation $\vec{\sigma}$ in the normal state (i.e. $g=0$) for the 
wire (a), ring (c) and elliptical ring in (e) and (g). The configuration in (c) and (d) are associated with the electron spin having a winding around the ${\cal{N}}$ and ${\cal{N}}$-$z$ directions. The superconducting ${\vec{d}}$-vector profiles in (e),(g),(g) and (h) are those corresponding to the spin patterns in (a),(b), (c) and (d), respectively. 
For convenience and clarity of graphical representation we employed different sets of microscopic parameters. In the superconducting state we have $g=0.5$ and $\mu=0.5$ for all the geometrical configurations, i.e. (e)-(h). The amplitude of the RSOC is $\alpha_{SO}$=0.053 for the wire and the ring. For the elliptical ring the RSOC amplitude is $\alpha_{SO}$=0.085 in (c) and (g), while $\alpha_{SO}$=0.12 is chosen for (d) and (h). The quantum configurations for the elliptical ring correspond with semi-axes ratio $a/b=0.3$. All the energy scales are in unit of 2$t$.}  
\label{fig:fig3}
\end{figure*}
{\it{Inhomogeneous spatial profile of the spin-singlet OP --}}
The elliptically deformed ring [Fig. 1 (d)] is an ideal platform to explore the consequences of the curvature on superconductivity because the imbalance of the semi-axis lengths naturally introduces two different regions with a larger (smaller) curvature strength as compared to a circular ring [Fig. 1 (c)]. 
Although the averaged OP is insensitive to the shape variation, an inspection of the local OP indicates an intricate interrelation among its variation, the spatial dependent curvature, and the RSOC [Fig. 3]. 
As a consequence of the Rashba field bending, indeed, the OP gets enhanced or suppressed in the region where the curvature is maximal and strongly inhomogeneous (e.g. nearby $K_{max}$ in Fig. 1 (d)), and the strength of the variation can be controlled through the elliptical semi-axes ratio $a/b$ [Fig. 3(a)] and the RSOC [Fig. 3(b)]. A distinct feature of this effect manifests in the character of OP modification. We find that the OP amplitude depends on the electron density in the nanostructure [Fig. 3 (c)], namely it increases close to half-filling while it gets suppressed or enhanced in the low (high) density regime when the chemical potential is close to the band edge. There are two main possible sources for the curvature-induced OP amplitude variation. First, a spatial modification of the local Rashba field is generally expected to induce a local conversion of spin-singlet into spin-triplet pairs that would result into a reduction of the spin-singlet OP when comparing the uniform with the inhomogeneous curvature profiles.
However, the inspection of the spatial spin-triplet correlations indicates that their amplitude is not much affected by a change in the electron density, thus confirming that the rate of conversion is weakly dependent on the filling. Therefore, such mechanism can be ruled out as a primary origin of the amplitude variation of the spin-singlet OP. 
On the other hand, a closer view of the local density of states (LDOS) in the normal state reveals that a change of the energy spectrum close to the Fermi level affects the strength of the superconducting OP. In order to intuitively understand how the curvature variation affects the LDOS one can observe that, given the energy dispersion $\epsilon(R)$ \cite{Gentile2013}
of electrons propagating along the quantum ring with constant radius $R$ and curvature $K=1/R$, a change $\delta R$ will imply a modification of the spectrum in the leading order of $\epsilon(R+\delta R) \sim \epsilon(R) (1-\frac{\delta R}{R})$. Hence, a  reduction (growth) of the ring's radius (curvature) corresponds to an effective shrinking of the bandwidth. This is, indeed, what is obtained when comparing the computed spatial dependent DOS along the elliptical loop at positions of maximum ($K_{max}$) and minimum ($K_{min}$) curvature, respectively, as demonstrated in Fig. 3 (d). Since the effective bandwidth reduction is energy independent, we find that the LDOS at $K_{max}$ is always enhanced with respect to $K_{min}$ except when the Fermi level is close to the band edge. The consequence on the superconducting amplitude follows the expectation of the BCS theory that a larger DOS at the Fermi level yields a stronger OP amplitude. We point out that the effect of tailoring the OP is general and goes beyond the case of the elliptical ring because it is mainly driven by the local strength of the curvature.

{\it{Spatial texture of the spin-triplet pairing amplitude --}} Since the bending of the RSOC nanostructure can torque the spin orientation by inducing windings around the local binormal ($z$) and normal ($\hat{\cal{N}}$) symmetry directions \cite{Ying2016}, the superconducting state is prone to exhibit spin-triplet correlations with non-trivial textures in space. For the present analysis, it is convenient to introduce the spin-triplet correlator in real space in terms of a bond ${\vec{d}}({s_i})$-vector components along the symmetry axes of the elliptical ring as, 
$d_{x}(s_i)=\frac{1}{2}(-\langle c_{\uparrow}(s_i) c_{\uparrow}(s_i+a_0) \rangle + \langle c_{\downarrow}(i) c_{\downarrow}(i+a_0) \rangle)$, $d_{y }(s_i)=\frac{i}{2}(\langle c_{\uparrow}(s_i) c_{\uparrow}(s_i+a_0) \rangle + \langle c_{\downarrow}(s_i) c_{\downarrow}(s_i+a_0) \rangle)$, and $d_{z}({s_i})=(\langle c_{\uparrow}(s_i) c_{\downarrow}(s_i+a_0) + \langle c_{\downarrow}(s_i) c_{\uparrow}(s_i+a_0) \rangle )$, with $s_i$ labeling the $i^{th}$ site of the curved nanowire and $a_0$ the inter-atomic distance. 
In uniform non-centrosymmetric superconductors or for surface superconductivity, the RSOC marks the lack of inversion center through a vector ${\vec{l}}_{{k}}$ that is odd under inversion \cite{Frigeri2004}. There, the spin-triplet pairing is not excluded \cite{Frigeri2004} and, indeed, an optimal configuration with ${{\vec{d}}}_{k}|| {\vec{l}}_{k}$ is expected to occur \cite{Frigeri2004}. 
Here, we deal with a spatial variation of the Rashba field when moving from the wire to the ring or the elliptical ring, and we focus on the regime where the RSOC is larger than superconducting gap. As expected, for the uniform wire the $\vec{d}$-vector lies in the plane of the one-dimensional system [Fig. 4 (e)], and is collinear to both the normal direction $\cal{N}$ [Fig. 1 (a)] and the electron spin orientation associated to one of the Kramers degenerate state at the Fermi level [Fig. 4 (a)]. A ring with a constant curvature introduces a non trivial $z$ component both in the single electron spin orientation [Fig. 4 (b)] and in the ${{\vec{d}}}$-vector pattern [Fig. 4 (f)] due to the spin-torque induced by the geometric curvature. Although the curvature is constant and uniform, the behavior of the spin of the electron pairs is qualitatively different from the chain as they acquire an out-of-plane $z$ component and the ${{\vec{d}}}$-vector is not collinear to the electron spin orientation. For the elliptical ring, the inhomogeneities in the RSOC field allows for a large variety of profiles. We find that the ${{\vec{d}}}$-vector can generally exhibit a three-dimensional pattern that is modulated in amplitude and orientation when moving from region with large to small curvature [Figs. 4 (g),(h)]. Different types of complex spatial profiles occur depending on the character of the electron spin pattern in the normal state. Indeed, when ${{\vec{\sigma}_z}}$ and ${{\vec{\sigma}_T}}$ change sign also the ${{\vec{d}}}$-vector manifests a similar variation [Figs. 4 (g),(h)]. Thus, remarkably, the $\vec{d}$-vector exhibits a complete winding around the normal direction $\hat{{\cal{N}}}$ [Figs. 4 (g),(h)] thereby following that of the electron spin orientation in the normal state [Figs. 4 (c),(d)]. Apart from the winding around the $z$ direction which is due to the loop geometry, the induced  ${{\vec{d}}}$-vector provides evidence for the occurrence of topologically non-trivial quantum states in all regimes of electron spin-texture. Interestingly, we find that the modulation in amplitude and orientation of the ${{\vec{d}}}$-vector [Fig. 4 (h)] is stronger when in the normal state the spin-texture has both windings around $z$, in the moving frame, and $\hat{{\cal{N}}}$ directions [Fig. 4 (d)].  

{\it{Conclusions --}} In conclusion, we demonstrate that the RSOC in shape-deformed nanostructures leads to novel paths for an all-geometric manipulation of the superconducting state, both for spin-singlet and spin-triplet quantum configurations. An immediate consequence of the observed effects is that the geometrically driven local change of the OP can be employed to engineer junctions with regions of enhanced or suppressed superconductivity integrated in a single material system and, in turn, to suitably tailor the transport properties.  
An experimental probe of the effects of the geometric curvature through the observation of the OP amplitude variation can be accessed by measuring the LDOS through scanning tunneling microscopy and comparing the gap in the spectrum at the positions of maximal and minimal curvature. 
We notice that the RSOC strength is large on the surface
states of high $Z$ metals, such as Au \cite{LaShell1996}, Bi \cite{Koroteev2004} and Pb \cite{Slomski2011}.
Therefore, feasible experimental platforms might be based on appropriately designed asymmetric nanorings of high $Z$ metals. Other paths to shape engineer the superconducting state could be the design of quasi-one-dimensional quantum wells at LAO-STO interface \cite{Ron2014,Bogorin2010,Stefhanos2012} or similar oxide heterostructures where the RSOC is strong, gate tunable and the shape design can be achieved by electrical means. 
Finally, on the basis of our results, we envision novel scenarios for the occurrence and geometric manipulation of topological states in heterostructures based on elliptically or asymmetric shaped semiconducting rings \cite{Fomin2014,Blossey2002,Offermans2005,Wu2012,Raz2003,Sormunen2005} as well as by employing suitably shaped topological insulators \cite{MolenkampHgTe} interfaced with $s$-wave superconductors. The emergence of non-trivial winding in space of the spin-triplet pairs also indicates the potential to get inhomogeneous topological states along curved nanostructures.

{\bf{Acknowledgements}}  We acknowledge the financial support of the Future and Emerging Technologies (FET) programme under FET-Open grant number: 618083 (CNTQC). 
CO acknowledges support from the Deutsche Forschungsgemeinschaft (grant No. OR 404/1-1) and from VIDI grant (project 680-47-543) financed by the Netherlands Organization for Scientific Research (NWO).

\end{document}